\documentclass[aip,graphicx]{revtex4-1}

\usepackage{graphicx}
\usepackage{dcolumn}
\usepackage{bm}
\draft 

\begin{document}


\title{Characterization of the HAYSTAC axion dark matter search cavity using microwave measurement and simulation techniques} 



\author{Nicholas M. Rapidis}
\affiliation{Physics Department, University of California, Berkeley, California 94720, USA}
\author{Samantha M. Lewis}
\email[]{smlewis@berkeley.edu}
\author{Karl A. van Bibber}
\affiliation{Nuclear Engineering Department, University of California, Berkeley, California 94720, USA}


\date{\today}

\begin{abstract}
Many searches for axion cold dark matter rely on the use of tunable electromagnetic resonators. Current detectors operate at or near microwave frequencies and use cylindrical cavities with cylindrical tuning rods. The cavity performance strongly impacts the signal power of the detector, which is expected to be very small even under optimal conditions. There is strong motivation to characterize these microwave cavities and improve their performance in order to maximize the achievable signal power. We present the results of a study characterizing the HAYSTAC (Haloscope At Yale Sensitive to Axion Cold dark matter) cavity using bead perturbation measurements and detailed 3D electromagnetic simulations. This is the first use of bead perturbation methods to characterize an axion haloscope cavity. In this study, we measured impacts of misalignments on the order of 0.001 in and demonstrated that the same impacts can be predicted using electromagnetic simulations. We also performed a detailed study of mode crossings and hybridization between the TM$_{010}$ mode used in operation and other cavity modes. This mixing limits the tuning range of the cavity that can be used during an axion search. By characterizing each mode crossing in detail, we show that some mode crossings are benign and are potentially still useful for data collection. The level of observed agreement between measurements and simulations demonstrates that finite element modeling can capture non-ideal cavity behavior and the impacts of very small imperfections. 3D electromagnetic simulations and bead perturbation measurements are standard tools in the microwave engineering community, but they have been underutilized in axion cavity design. This work demonstrates their potential to improve understanding of existing cavities and to optimize future designs.
\end{abstract}


\maketitle 

\section{Introduction}\label{sec:intro}
Modern astrophysical and cosmological observations have shown that only a small fraction of the universe is composed of ordinary matter. Recent measurements predict that $\sim26\%$ of mass-energy in the universe is dark matter.\cite{PlanckResult} The makeup of the dark matter remains unknown, though several candidates have been proposed. Among them is the axion, a light hypothetical pseudoscalar particle.\cite{PhysRevLett.38.1440,PhysRevD.16.1791,PhysRevLett.40.223,PhysRevLett.40.279,PRESKILL1983127}

Detecting a dark matter particle is inherently challenging because the dark matter has extremely weak interactions with ordinary matter. Axion detection schemes are largely based on an axion--photon coupling interaction known as Primakoff conversion.\cite{PhysRevLett.51.1415} In the presence of a strong magnetic field, the axion scatters off of a virtual photon and converts to a real photon that can be detected. The frequency of the resulting photon corresponds directly to the axion mass as
\begin{equation}
\nu_a = \frac{m_ac^2}{h}\,,
\label{eq:nua}
\end{equation}
where $m_a$ is the axion mass, $c$ is the speed of light, and $h$ is the Planck constant. Both $m_a$ and the coupling strength of the axion--photon conversion $g_{a\gamma\gamma}$ are unknown, but the parameter space is bounded by astrophysical constraints and observations.\cite{Graham} Axion searches aim to scan the possible parameter space for a signal and eliminate regions where no signal is detected. The axion--photon coupling is expected to be very weak and therefore the power of an axion photon signal would be very small (on the order of $10^{-24}$ W).\cite{ALKENANY201711}

\begin{figure}
\includegraphics[width=\columnwidth]{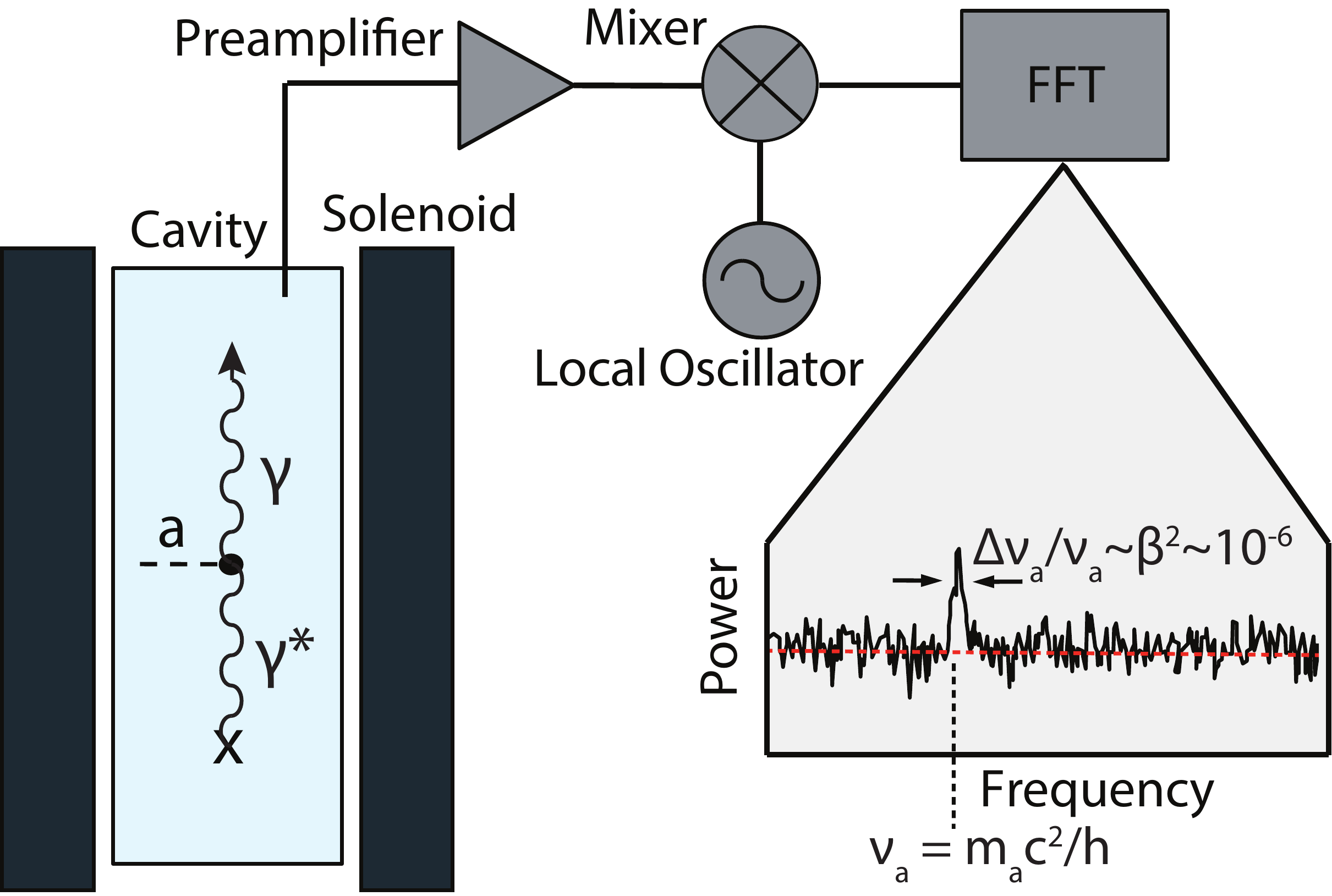}
\caption{Block diagram of a haloscope detector. A resonant cavity is placed inside the bore of a superconducting solenoid. An axion in the cavity converts to a photon via Primakoff effect. This signal is extracted and amplified. The spectrum in the bottom right corner shows the expected shape of an axion signal, which appears as a power excess above the background. The shape of the axion results from its virial velocity in the galactic halo. \label{fig:blockdiagram}}
\end{figure}

Existing searches for galactic halo axions use detectors known as haloscopes. The Haloscope At Yale Sensitive To Axion CDM (HAYSTAC) is a haloscope searching specifically for Cold Dark Matter (CDM) axions, which are assumed to have a uniform `cold' virial velocity. In these devices, a tunable electromagnetic cavity is placed in the bore of a solenoid. The goal is to match the frequency of a resonant cavity mode of choice to $\nu_a$ in order to enhance the signal from the converted photon. An antenna is used to extract the signal, which is then amplified and analyzed. A photon resulting from inverse Primakoff conversion would appear as a power excess. A block diagram of this detection technique is shown in Figure \ref{fig:blockdiagram}. Haloscopes operate at cryogenic temperatures and use electronics operating at or near the Standard Quantum Limit for noise in order to be sensitive to $P\sim10^{-24}$ W.

The haloscope signal power scales as
\begin{equation}
P \propto B_0^2VQ_LC_{nm\ell}g^2_{a\gamma\gamma}\,,
\label{eq:power}
\end{equation}
where $B_0$ is the magnetic field supplied by the solenoid, $V$ is the cavity volume, and $Q_L$ is the loaded cavity quality factor. $C_{nm\ell}$ is the `form factor' of the resonant cavity mode, defined as
\begin{equation}
C_{nm\ell} \equiv \frac{\left(\int{d^3\bm{x}\bm{\hat{z}}\cdot\bm{E}_{nm\ell}\left(\bm{x}\right)}\right)^2}{V\int{d^3\bm{x}\varepsilon\left(\bm{x}\right)\left|\bm{E}_{nm\ell}\right|^2}}\,.
\label{eq:formfactor}
\end{equation}
Here, $\bm{E}_{nm\ell}$ is the electric field of the mode, $V$ is the cavity volume, and $\varepsilon$ is the relative permittivity. Since the applied magnetic field is in the $z$ direction, the form factor quantifies the overlap between the external applied magnetic field and the electric field of the mode.

Current experiments use cylindrical microwave cavities with cylindrical metal tuning rods. These cavities are simple from a resonator design perspective, but use in axion haloscopes poses unique constraints and challenges. The speed of axion searches is limited by the haloscope signal power, which in turn depends heavily on cavity performance. In designing the first axion haloscope cavities, Hagmann \textit{et al.}\cite{Hagmann} recognized the importance of studying and limiting non-ideal cavity behavior. Their analysis resulted in the development of the haloscope cavities that are in use today.

Now, modern computational tools are capable of modeling these cavities in much greater detail. Coupling these simulations with measurement techniques used in other cavity applications, it is possible to study the imperfections of existing cavities and develop a predictive capability for new designs. This paper outlines the use of standard microwave measurement techniques and simulation software to fully characterize the cavity for the HAYSTAC detector. This is the first use of the bead perturbation technique to study an axion haloscope cavity and the first attempt to match these measurements with 3D electromagnetic simulations of the cavity with imperfections. 

\section{HAYSTAC}\label{sec:HAYSTAC}
HAYSTAC is a collaboration of Yale University, the University of Colorado Boulder, and the University of California Berkeley. Housed at Yale, the detector uses a 9 T superconducting magnet and is cooled to 127 mK using a dilution refrigerator. Josephson Parametric Amplifiers (JPAs) are used to amplify the microwave signal with quantum-limited added noise. With the cavity described in Section \ref{subsec:cavity}, HAYSTAC has excluded axion masses of 23.15--24.0 $\mu\text{eV}$ within the axion dark matter model band.\cite{PhysRevLett.118.061302,PhysRevD.97.092001}

\subsection{Cavity design}\label{subsec:cavity}
The current HAYSTAC cavity consists of a right cylinder barrel with two endcaps and a large (radius $r = 1$ in) cylindrical tuning rod. The barrel has an inner radius $R = 4$ in and length $L = 10$ in. The barrel and endcaps are constructed from stainless steel and plated with copper for good electrical and thermal conductivity at cryogenic temperatures. Measurements for this study were performed at room temperature using an aluminum twin of the main cavity barrel, as the main cavity was in use for data collection. In both the main cavity and the twin cavity used for this study, the tuning rod is hollow with a copper-plated stainless steel body.

The tuning rod has an off-axis axle composed of a steel spindle inside the body of the rod and an alumina tube on each end (outer diameter $d = 0.25$ in). The rod is slightly shorter than 10 in, leaving small ($\sim 0.01$ in) gaps between the rod and the endcaps. While these gaps are necessary in order for the rod to rotate, they have a substantial impact on the cavity performance. This is discussed in more detail in Section \ref{subsec:future}. Turrets are attached to the endcaps to hold the alumina portions of the tuning rod axle. A ball bearing is the only point of contact between the axle and the turret, allowing for frictionless rotation.

\begin{figure}
\includegraphics[width=\columnwidth]{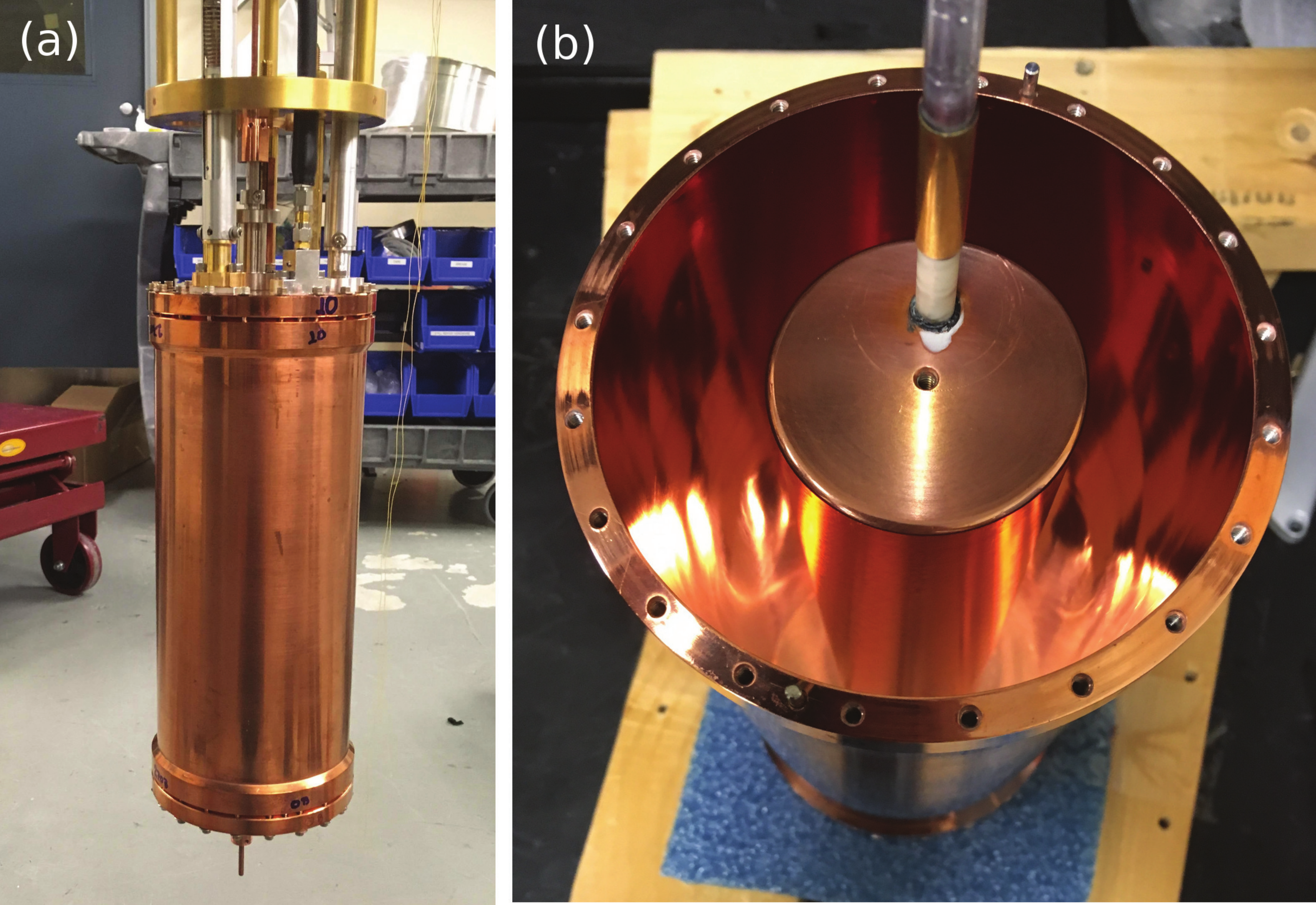}
\caption{(a) Photograph of the assembled HAYSTAC cavity before being inserted into the magnet. (b) The HAYSTAC cavity with the top endcap removed, showing the tuning rod. Both the cavity and the rod are stainless steel plated with copper. A portion of the top axle is also visible. \label{fig:cavityphotos}}
\end{figure}

\begin{figure}
\includegraphics[width=\columnwidth]{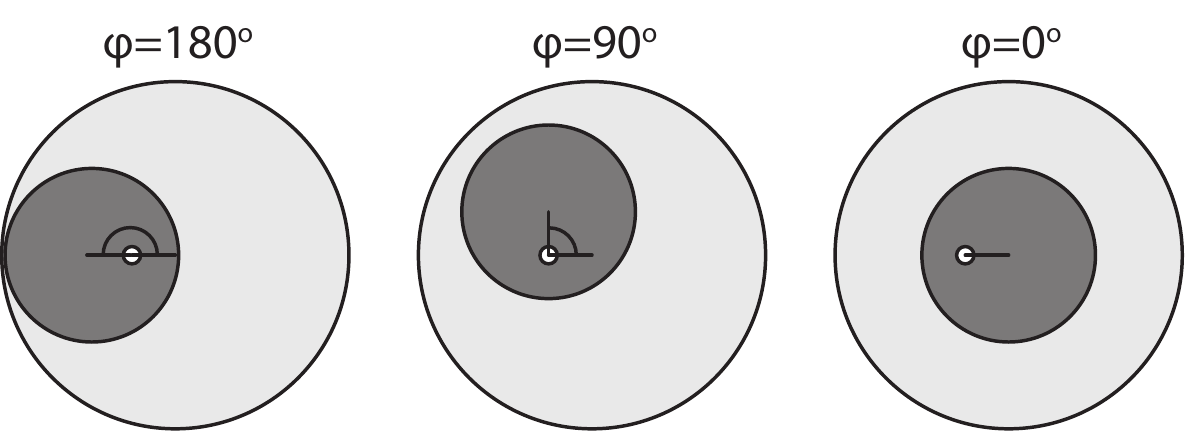}
\caption{Top-down schematic of the cavity (light gray) and rod (dark gray) showing the angle definitions for different rod positions. \label{fig:rodangles}}
\end{figure}

During HAYSTAC data collection, the cavity mode of interest is the lowest order transverse magnetic mode, the TM$_{010}$. This mode has an electric field purely along the cavity axis, giving it the highest possible form factor of any fundamental cavity mode. To tune the cavity, the rod is rotated from being concentric with the cavity body ($0^{\circ}$, the maximum frequency position) to being nearly tangent with the wall ($180^{\circ}$, the minimum frequency position). Photographs of the cavity and tuning rod are shown in Figure \ref{fig:cavityphotos}; Figure \ref{fig:rodangles} shows a schematic of the tuning extrema. This 180$^{\circ}$ rotation tunes the TM$_{010}$ mode between 3.4--5.8 GHz.

\subsection{Cavity challenges and limitations}\label{subsec:cavchallenges}
Right cylindrical cavities are well-suited for axion haloscope experiments at microwave frequencies. However, there are challenges in using even simple cavity designs. Initial scans have been performed in narrow regions of the cavity tuning range, but as the community moves to cover more parameter space it will be necessary to tune through regions where performance is limited. The cavity performance impacts the signal power through three parameters: the volume, the quality factor, and the form factor. It is therefore vital to understand the cavity performance in problematic regions and make improvements wherever possible.

\subsubsection{Mode crossings}\label{subsubsec:probsmodecross}
Microwave resonators have many fundamental eigenmodes, but few of these modes have large enough $C_{nm\ell}$ values to be useful in current haloscopes because their $E_z$ components are small or zero. When the tuning rod is rotated, all of the TM (transverse magnetic) eigenmodes tune. The transverse electric (TE) and transverse electromagnetic (TEM) modes present in the cavity are unaffected by this tuning motion. Thus as the frequency of the TM$_{010}$ mode is tuned, the mode crosses stationary TE and TEM modes (generically referred to as `intruder modes'). In an ideal cavity there would be no interaction between the modes, but in reality they mix and form hybrid modes. This degrades the $E_z$ field strength and lowers $C_{010}$. Data taken in these regions is effectively unusable for excluding axion parameter space.

To date, HAYSTAC and other haloscopes have circumvented this problem by focusing on frequency ranges that are free from mode crossings and by removing any compromised portions of the data from new exclusion limits. As HAYSTAC moves to perform scans over broader frequency ranges and at higher frequencies, this solution is no longer tenable. Given the density of modes in the HAYSTAC cavity, there are few regions that are entirely free from mode crossings. A broad sweep will therefore be broken up by several of these unusable regions. The density of modes is also higher at higher frequencies, making it difficult to find substantial crossing-free regions. Understanding the nature of mode crossings and characterizing the range over which mixing occurs may allow for useful data to be obtained over part or all of the crossing. A major goal of this study was to characterize the mode crossings in the HAYSTAC cavity such that portions of these regions may be used despite the mode hybridization.

\subsubsection{Alignment and mode localization}\label{subsubsec:probslocalization}
In the ideal case, the TM$_{010}$ electric field strength is constant along the length of the cavity. The entire field strength is in the $z$ component of the field except at boundaries, which gives the TM$_{010}$ its high form factor. Deviations from perfect alignment disrupt this ideal behavior and cause field strength localization. Notably, tilting the tuning rod with respect to the cavity barrel causes an asymmetry in the axial direction, which forces the field to localize towards one end of the cavity.

The impact of this mode localization on the cavity parameters $Q_L$ and $C_{010}$ has been largely unstudied, and power calculations have generally relied on idealized $C_{010}$ values. This work details efforts to characterize and quantify the effects of small tilts on these parameters through measurement and simulation, allowing for improved accuracy in power calculations. This study also aims to demonstrate the predictive capability of simulations, which can be used to inform future cavity designs.

\section{Measurement and simulation techniques}\label{sec:techniques}
\subsection{Bead perturbation}\label{subsec:bead}
In order to study mode localization, it is necessary to determine the relative field strength at different locations inside the cavity. To perform a bead perturbation measurement, a small metal or dielectric object is inserted into a resonator to locally perturb the electric field and cause a shift in the resonant frequency of an eigenmode. By performing this measurement at different locations, a `map' of the field strength can be constructed. This technique is commonly used in accelerator physics to study the field profile of a cavity or set of cavities along the axis of acceleration.\cite{BP1,BP2,BP3}

For axion haloscope cavities, a series of bead perturbation measurements can be taken along $z$ at one radial and azimuthal position (referred to in this work as a `bead-pull' measurement). Since the TM$_{010}$ mode has a constant $E\left(z\right)$ for any given $r$ and $\theta$, the result should be a profile with a single, constant value for the perturbed frequency at each axial position. Any deviations from this flat profile indicate there is mode localization causing the field to be stronger at some location. The details of how this measurement technique was used to characterize the HAYSTAC cavity are discussed in Section \ref{subsec:setup}.

\subsection{Electromagnetic simulations}\label{subsec:fem}
Electromagnetic cavities can be simulated in 3D using finite element modeling. The finite element method (FEM) is used across many disciplines to solve systems of equations in complex geometries. In FEM problems, a geometry is broken up into discrete elements over which the system of equations is easier to solve. Together, the elements form a mesh. Numerical solutions are found for each element and are used to construct a solution for the entire geometry. This can be applied to full 3D systems with many constituent materials and components of different sizes.

The use of 3D electromagnetic (EM) simulations allows for the design of cavities and other devices which cannot be described analytically. By solving Maxwell's Equations, 3D EM simulations can predict the eigenmodes of electromagnetic resonators. Material information can be included to determine expected $Q$ values and losses. When excitation sources such as antennas or waveguide ports are included, 3D EM simulations can determine the expected transmission and reflection behavior of a structure. All of this information is used in the design of microwave cavities and can be compared to real values measured in a structure. In this study, simulations were performed using a commercial electromagnetic simulation software, Computer Simulation Technology Microwave Studio (CST MWS).\cite{CST}

\section{Methodology and results}\label{sec:methodology}

\subsection{Measurement setup}\label{subsec:setup}
The bead-pull measurement setup has three main components: the beadline and a stepper motor system, a vector network analyzer (VNA), and a LabVIEW script that controls the VNA and the motor system.

The beadline consists of a cylindrical sapphire bead of length 0.196 in and diameter 0.169 in that is glued onto a Kevlar\cite{Kevlar} string. The string passes through a hole in the endcaps, situated at $r=1.25$ in (with respect to the center of the cavity). When properly aligned, the bead can traverse the cavity at a constant radial and azimuthal position.

Motion of the beadline is controlled by a system of pulleys and a stepper motor. The pulleys are adjusted such that the beadline is parallel to the z-axis of the cavity. The Kevlar line is attached to an Applied Motion Products STM 23S-2EE stepper motor. The motor rotates in predetermined steps which lead to changes in the vertical position of the sapphire bead.

A second stepper motor is attached to the tuning rod axle, providing fine control of the rod's rotation for tuning. The cavity and motor are attached to a frame separate from the beadline system. Vibration isolation is included to prevent the motor from inducing vibrations in the cavity.

Two antennas made from semi-rigid coax are inserted through the bottom endcap of the cavity. A Keysight E5071C Network Analyzer provides information about the transmission and reflection measurements between the two antenna ports throughout the frequency range of the cavity. Transmission measurements $\left(S_{21}\right)$ are used to determine the frequencies and $Q$ values of the cavity modes, which appear as peaks on a spectrum of power versus frequency.

The stepper motors and the VNA are controlled by a LabVIEW program. For each measurement, the user sets the desired number of bead steps and the frequency of the mode of interest. The rest of the process is automated: the bead is moved and the software records the central frequency of the peak in $S_{21}$. After all steps are performed, the software plots the central frequency versus bead position. This `profile' of the mode is used to determine if there is mode localization or hybridization.

\subsection{Tilt study}\label{subsec:tiltstudy}
\begin{figure}
\includegraphics[width=\columnwidth]{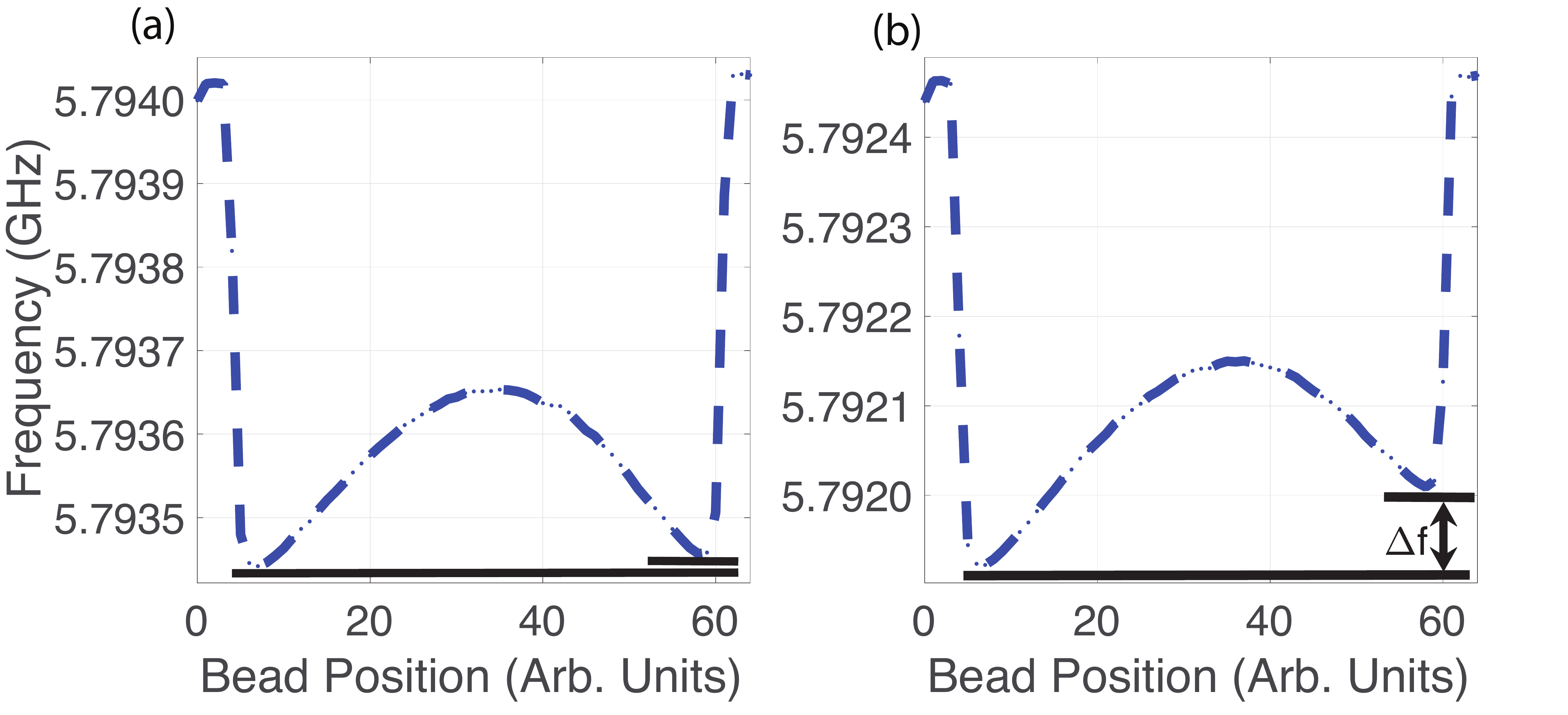}
\caption{Two profiles of the TM$_{010}$ mode produced using bead-pull measurements. The frequency of the mode shows a dependence on the position of the bead in the cavity. The abrupt changes in frequency on the two ends correspond to the points where the bead enters and exits the cavity. An approximately flat bead-pull profile (a) is nearly symmetric about the cavity center. Profiles for the misaligned case (b)  exhibit a non-negligible $\Delta f$. In this case the $\Delta f = -90\ \textrm{kHz}$. The curved shape of the profile arises from mode localization in the gaps between the endcaps and tuning rod. This is discussed in further detail in Section \ref{subsec:comparison}. \label{fig:deltaf}}
\end{figure}

Misalignments of the rod with respect to the cavity cause mode localization that can generally be observed through bead-pull measurements. This mode localization manifests as a shift in the resonant frequency of the TM$_{010}$ mode from one end of the cavity to the other. A primary goal of this study was to quantify this shift and relate it to a physical amount of misalignment while studying its impact on cavity performance. The frequency shift was defined as 
\begin{equation}
\Delta f \equiv f\left(z=\text{cavity bottom}\right)-f\left(z=\text{cavity top}\right)\,.
\label{eq:delta}
\end{equation}
Non-zero $\Delta f$ indicates there is some mode localization. Figure \ref{fig:deltaf} shows two example bead-pull profiles with different $\Delta f$ values.

\begin{figure}
\includegraphics[width=\columnwidth]{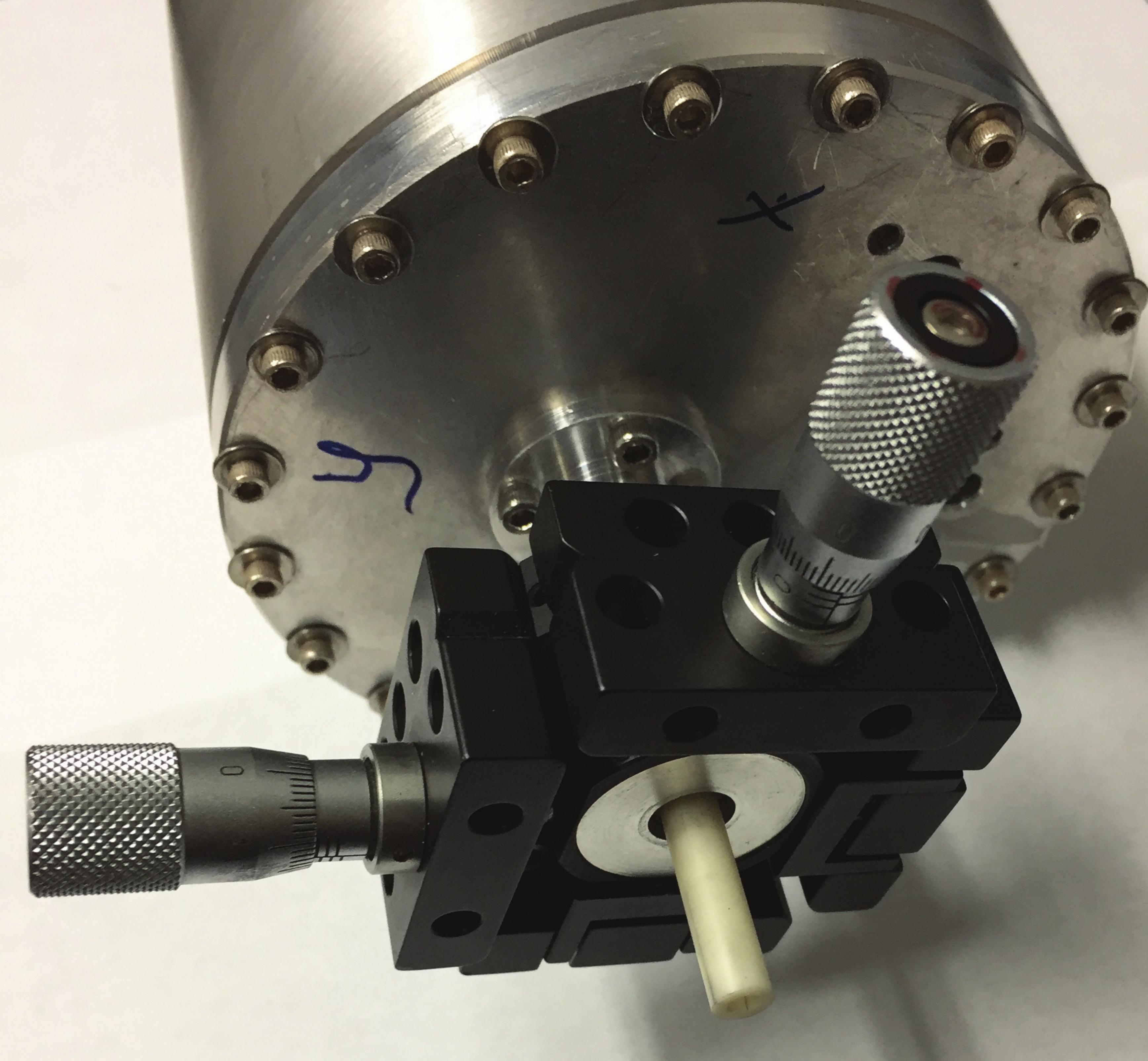}
\caption{Micrometer system used to control tilt of rod. The tight collar is not shown in this photo.\label{fig:microm}}
\end{figure}

The rod misalignment was produced by laterally shifting the rod's axle at one end of the cavity. Two orthogonal micrometers, as shown in Figure \ref{fig:microm}, were mounted on the turret on the bottom endcap of the cavity around the alumina axle. By adjusting the micrometers, the rod can be tilted in a controlled and measurable manner. Rod tilt studies were conducted at three approximate rod positions $\varphi$: 0$^\circ$, 90$^\circ$, and 180$^\circ$, which correspond to the positions illustrated in Figure \ref{fig:rodangles}. These positions were chosen because they covered the extremal frequency cases and a central frequency. Once the rod was adjusted to a position of interest, a tight collar was placed on the top alumina axle to hinder rotations of the rod within the cavity. This reduced the degrees of freedom of the system to the two orthogonal tilts of the rod. The micrometers were set such that the starting values corresponded to $\Delta f = 0$, taken to be an aligned configuration.

The tilt studies were carried out by adjusting the micrometers in the two orthogonal directions and taking a bead-pull measurement at each position. A square grid of micrometer positions corresponding to rod misalignments was created. In the 0$^\circ$ and the 180$^\circ$ cases, the measurements were taken based off of a $9\times9$ square grid of 0.8 mil steps in the two orthogonal directions, i.e. the maximum micrometer displacements were $\pm4\times0.8\ \textrm{mils}=\pm3.2\ \textrm{mils}$. In the 90$^\circ$ case, a decrease in the mode's sensitivity to misalignments demanded an increase in the grid's dimensions in order to see an appreciable impact. This resulting grid was $7\times 7$ with a 1.2 mil step size between points. These measurements kept the angular misalignment between the rod and the cavity below 1 millirad.

\begin{figure}
\includegraphics[width=\columnwidth]{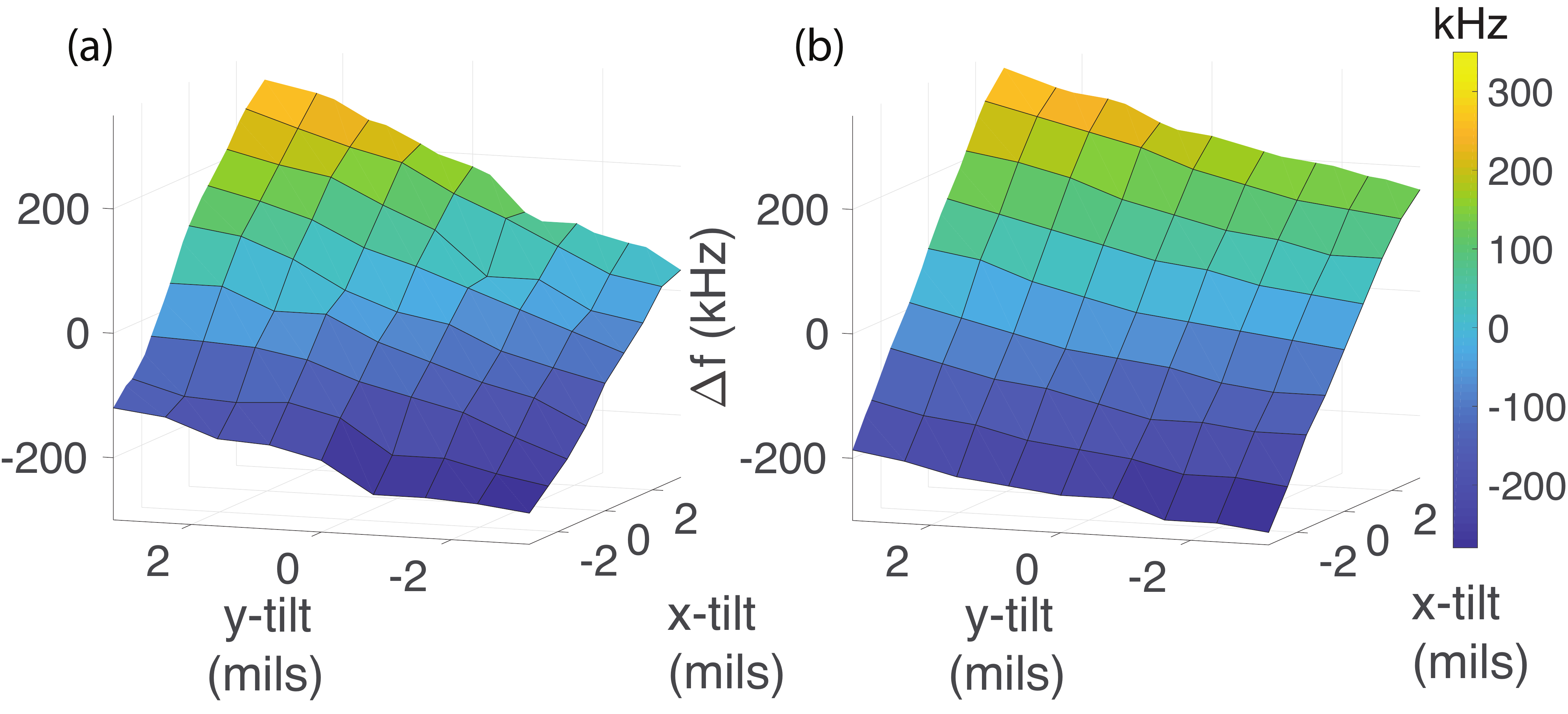}
\caption{Planes created by studying $\Delta f$ as a function of rod misalignment from (a) the measurements and (b) the simulations. The misalignment is depicted in terms of the step size of the micrometer. One step in each direction corresponds to an adjustment of the microemeter's position by 0.8 mils. These planes correspond to a rod rotation of $\varphi=3.3^{\circ}$. \label{fig:tiltplanes}}
\end{figure}

The behavior of the cavity at the three rod positions resulted in a plane-like surface in $\Delta f$, as shown in Figure \ref{fig:tiltplanes}a. The orientation of the set of points along which $\Delta f \approx 0$ showed a dependence based on the angle of rotation of the rod.

The same misalignments were input into a CST MWS model of the cavity. Bead-pull profiles were constructed by simulating the measurement process: a sapphire bead was stepped through the cavity and the frequency of the TM$_{010}$ mode was determined from $S_{21}$. The simulated bead-pulls were designed such that they matched the conditions of the measured bead-pulls. Stepping of the bead was performed using a parametric sweep based on the bead's position with mesh adaptation between steps. These simulated bead-pulls were performed at each tilt position in the measurement grids. Figure \ref{fig:tiltplanes}b shows the resulting plane in $\Delta f$.

Additional eigenmode simulations were performed with various tilts to study the impact of tilt and mode localization on the form factor $C_{010}$. It has previously been assumed that any mode localization would cause a reduction in $C_{010}$. However, simulations showed that small amounts of axial mode localization introduced by the tilts had negligible impacts on $C_{010}$ because the overall $E_z$ is conserved. Under ideal conditions, the form factor is largest when the field configuration is most symmetric at $\varphi = 0^{\circ}$, decreases through $\varphi = 90^{\circ}$, and increases near $\varphi = 180^{\circ}$. Simulations at each of these three positions showed that, for a given tilt, the impact is largest when the form factor is highest. Tilts in the range of 4--5 mils resulted in decreases in the form factor of 2.4\%, 0.5\%, and 1.5\% for $\varphi = 0^{\circ},~90^{\circ},~\text{and}~180^{\circ}$, respectively. For smaller tilts in the range of 1--4 mils, the shift in form factor was generally less than 0.5\%.

\subsection{Mode crossings study}\label{subsec:modestudy}
\begin{figure*}
\centering
\includegraphics[width=\textwidth]{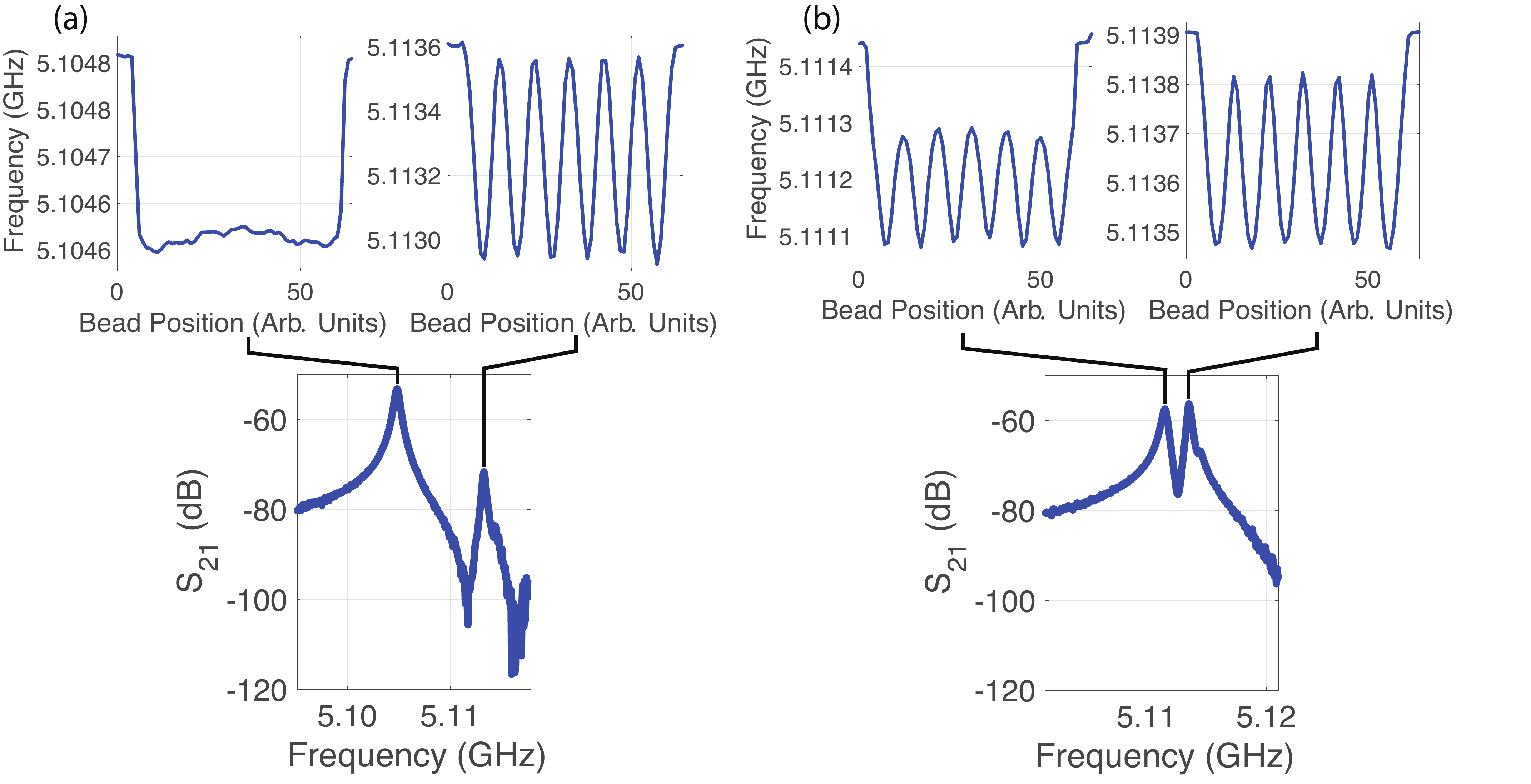}
\caption{Examples of bead-pull profiles near the edge and the middle of a mode crossing. (a) Shows the very beginning of a mode crossing. This is identified by the introduction of small oscillations in the TM$_{010}$ mode profile (left). (b) As the modes approach in frequency, they hybridize and their profiles superimpose. At this point, neither mode can be used for data acquisition. The bead-pull profile as well as 3D EM simulations illustrate that the mode with the oscillatory profile is the TE$_{060}$ mode. \label{fig:modecrossing}}
\end{figure*}

The primary goal of the mode crossings study was to determine the width of mode crossings between the TM$_{010}$ mode and other TE and TEM modes in the cavity. The width of the mode crossing was defined to be the frequency range across which the two modes were sufficiently hybridized such that the effects of this hybridization were visible on a bead-pull profile. Effects of hybridization were observed on bead-pull profiles when the profile of a TE or TEM mode was superimposed on the profile of the TM$_{010}$ mode, as shown in Figure \ref{fig:modecrossing}. The regions in which the modes mix are used to determine the frequency ranges over which the TM$_{010}$ mode can no longer be used due to its hybridization.

This study was conducted by performing a series of bead-pull measurements with small rotations of the tuning rod. The rod position $\varphi$ was controlled by a stepper motor so very small steps could be made. Bead-pull measurements were taken every 5 MHz in frequency throughout the entire tuning range in order to look for mode crossings. When a mode crossing was encountered, the rod tilt was adjusted to give $\Delta f = 0$ on the TM$_{010}$ mode. The rod was then rotated in steps corresponding to 2.5 MHz in frequency to bring the frequency of the TM$_{010}$ mode closer to that of the intruder mode. This was repeated until the TM$_{010}$ mode emerged at a frequency higher than that of the intruder mode and mixing was no longer observed. During each step in rotation, bead-pull measurements were taken on both peaks in $S_{21}$, producing two profiles. The alignment of the rod was not adjusted throughout the mode crossing as a control. The frequency range of mode crossings was sufficiently small such that the change in $\Delta f$ was negligible across the range of rotation.

These studies showed that approximately 15\% of the available frequency range of the TM$_{010}$ mode contains significant mode mixing that must be considered during axion search data collection. Several observed mode crossings proved to be fairly innocuous since no mixing was observed unless the mode with the wider and more dominant peak (in all such cases, the TM$_{010}$ mode) completely overlapped with the intruder mode. 

\begin{figure}
\includegraphics[width=\columnwidth]{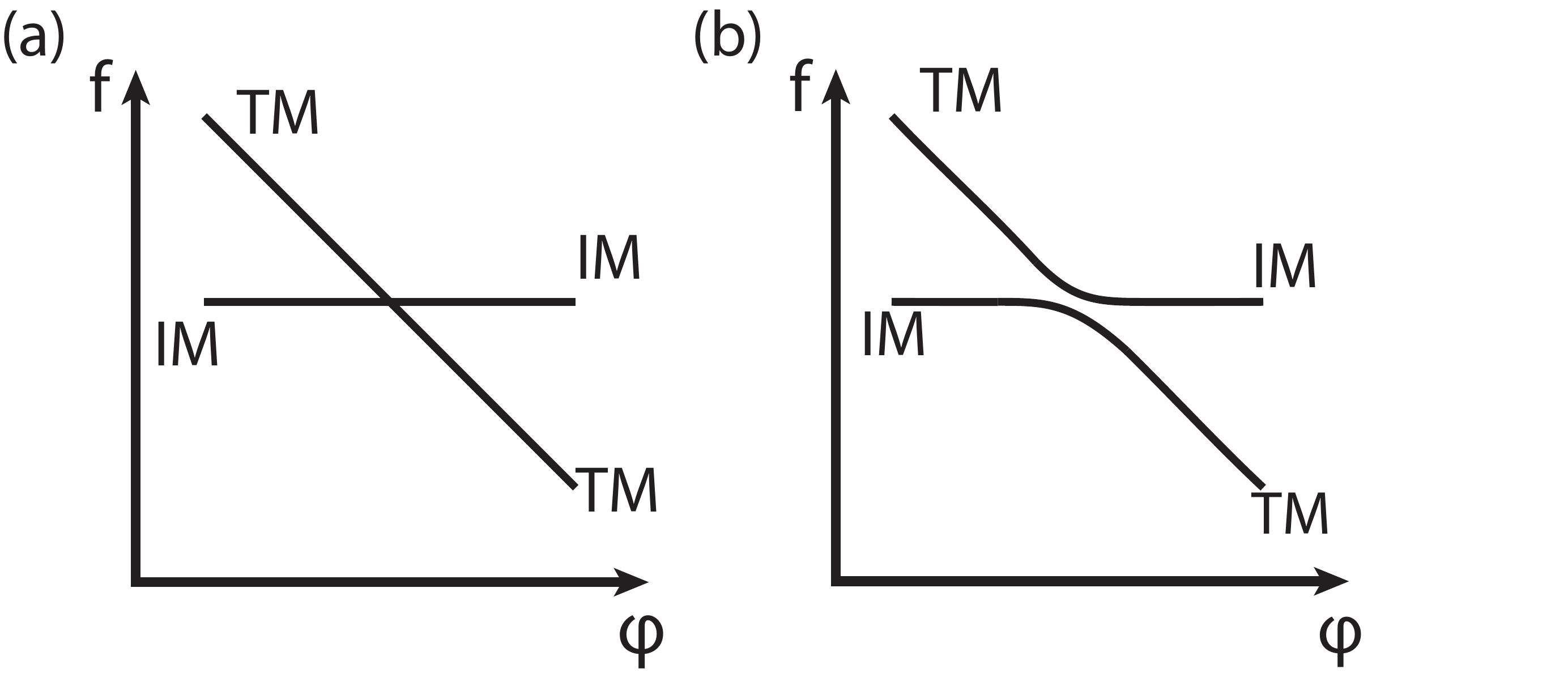}
\caption{Two types of mode crossings between the TM modes and the intruder modes (IM) in which (a) the intruder mode's frequency remained constant throughout the crossing while the TM mode tuned through it and (b) an `avoided crossing' takes place. \label{fig:avoidedcrossings}}
\end{figure}

Stronger and wider mode crossings exhibited one of two behaviors. In some cases, the frequency of one hybridized mode remained constant while the other mode tuned across the frequency range of interest. The stationary hybridized mode corresponded to the peak which was initially the intruder mode. This type of crossing is illustrated in Figure \ref{fig:avoidedcrossings}a. In the remaining cases, the mode that initially was at the lower frequency tuned to the original frequency of the intruder mode, while the intruder mode tuned away and eventually became the TM$_{010}$ mode. In these cases, the two peaks never truly crossed, making them so-called `avoided crossings,' as shown in Figure \ref{fig:avoidedcrossings}b. 

Mode crossing simulations were performed in CST MWS. To model a mode crossing, bead-pulls were simulated as described in Section \ref{subsec:tiltstudy} at multiple rod angles. Profiles were constructed from both the TM$_{010}$ mode and the intruder mode to be compared directly with the measured results. Simulated mode crossings agreed with measured results for the width and behavior of the mode crossing.

Additional mode crossing simulations were performed using the properties of copper at cryogenic temperatures. While the $Q$ value increases by roughly a factor of 4 when the cavity is cooled, there is negligible change in the width of mode crossings. For example, the mode crossing shown in Figure \ref{fig:modecrossing} exhibits mixing over 15 MHz, corresponding to $0.4^{\circ}$ in rod rotation. Simulations were performed in steps of $0.05^{\circ}$, corresponding to roughly 1 MHz in frequency. On this order, the widths of crossings were the same at room temperature and cryogenic temperatures.

Simulating mode crossings provides additional information which cannot be determined via measurements, including whether the intruder mode is a TE or TEM mode. Bead-pull measurements show variations in the electric field strength along the axis, but do not give information on the orientation of the field which is necessary to distinguish the mode type. In simulations, the full 3D electric and magnetic fields can be calculated for any given frequency thus allowing each intruder mode to be characterized as TE or TEM.

\section{Discussion}\label{sec:discussion}

\subsection{Comparison of measured and simulated results}\label{subsec:comparison}
Across this study, measured and simulated results show good agreement. In all cases, the results are qualitatively the same. Small quantitative differences are found in many cases, which is expected given the limited complexity of the simulated model.

One method of illustrating the general agreement between measurements and simulations is to compare the frequencies of intruder modes in a given frequency range. In particular, in the 4--5 GHz frequency range, all measured intruder modes were observed in the simulations. As shown in Table \ref{table:intrudermodes}, their corresponding frequencies showed general agreement with measured values.

\begin{table}
\begin{tabular}{c c}
\toprule
\textbf{Simulated Frequencies (GHz)}        & \textbf{Measured Frequencies (GHz)} \\ \hline
4.135                                                                    & 4.136                                                         \\ \hline
\begin{tabular}[c]{@{}c@{}}4.3338\\ 4.335\\ 4.3999\\ 4.4006\end{tabular} & Four modes in 4.327-4.365 region                              \\ \hline
\begin{tabular}[c]{@{}c@{}}4.7264\\ 4.7435\end{tabular}                  & Two modes in 4.710-4.730 region                                \\ \hline
4.8257                                                                   & 4.820                                                         \\ \hline
\begin{tabular}[c]{@{}c@{}}4.8945\\ 4.902\end{tabular}                   & Two modes in 4.885-4.905 region                               \\
\botrule
\end{tabular}
\caption{Tabulated frequencies of the intruder modes in 4--5 GHz range as observed in 3D electromagnetic simulations and in measurements. This corresponds to a range in tuning rod angle $\varphi$ of $49^{\circ}$. \label{table:intrudermodes} }
\end{table}

Figure \ref{fig:tiltplanes} visually demonstrates that the measured and simulated tilt studies showed the same overall behavior of $\Delta f$ versus tilt. This agreement is found for all three angles where tilt studies were performed. The form factor simulations showed while that $\Delta f$ is a clear measure of a physical misalignment, does not have a perfect correspondence to a certain shift in $C_{nm\ell}$. The impact on $C_{nm\ell}$ has more dependence on whether the rod is tilted towards or away from the center of the cavity. This indicates that while mode localization occurs with tilts, it does not necessarily degrade the form factor. Instead, tilts have the most impact when they cause mode localization in an otherwise highly symmetric field. Thus, $\Delta f$ is best used to match a measured frequency shift to a physical tilt. This physical tilt can then be included in simulations to calculate the form factor.

The simulated $\Delta f$ values agree quantitatively with measured values, though there are small differences. There are multiple possible sources for the discrepancies. In the measurement case, the accuracy of the setup is limited. Most notably, very small rotations of the tuning rod are still possible even though the rod is `fixed.' As such, the simulated rod angle may be slightly different from the measurement case. The accuracy of the micrometers is also limited, as they are set manually.

There are further known non-idealities which were not included in simulations in order to limit the complexity of the problem. The cavity barrel is not perfectly cylindrical and has small variations both radially and axially. These variations are on the order of 0.005 in or smaller ($\leq 0.15\%$ of the cavity diameter), but it is possible that they impact mode localization. Similarly, the tuning rod has small variations in its radius on the same order. The simulated cavity barrel and tuning rod are perfect cylinders and thus do not capture these potential impacts. Such variations could potentially be included in future studies by creating a full 3D model of the fabricated cavity.

An important potential source of error is the size of the gaps between the endcaps and the tuning rod. The length of the rod and the cavity are both known to a high degree of accuracy, but there is no mechanism to ensure the gaps are symmetric on the top and the bottom. When the cavity is assembled, the collar is attached to the top rod axle fixing the axial position of the tuning rod. This position is set such that the rod is not touching either endcap, but it cannot be set with the accuracy necessary to enforce a 0.01 in gap on either side. Thus, it is possible for the gaps to be asymmetric and the value of that asymmetry is unknown.

Simulations have shown there is strong mode localization in the gaps which gives rise to the parabolic shape of the profile shown in Figure \ref{fig:deltaf}. The impact varies with rod angle, giving profiles with different concavities. Asymmetry in the gaps can lead to shifts in this bump that may cause errors in calculating $\Delta f$. The simulated model assumed symmetrical gaps and thus did not include this source of error. To achieve better quantitative agreement in $\Delta f$, this gap asymmetry could be studied in more detail.

All observed intruder modes were found in the simulations at the expected frequencies. Small differences in resonant frequency between simulations and measurements are expected due to the non-idealities of the cavity that have been discussed. Furthermore, the simulations at cryogenic conditions showed that despite an increase in $Q$, the width of the mode crossings showed no observable temperature dependence. Overall, the simulated room-temperature mode crossings agreed well with measured values, demonstrating that simulations can be used to characterize the severity of mode crossings even before a cavity is fabricated.

\subsection{Implications for HAYSTAC and future cavity designs}\label{subsec:future}
The purpose of this study was not only to characterize the HAYSTAC cavity using microwave measurement techniques, but also to demonstrate the capability of 3D electromagnetic simulations to replicate the measurements. This opens up the opportunity to obtain more accurate information on cavity performance for the existing HAYSTAC cavity as well as potential future designs.

While the $Q$ value of the TM$_{010}$ mode can be measured \textit{in situ} during data collection, the form factor $C_{010}$ can only be calculated from simulated values of the $E$ field. Previous calculations used an idealized cavity. This study demonstrates that commercial 3D EM software is capable of capturing the effects of small non-idealities such as rod misalignments. Future $C_{010}$ calculations can include misalignments and other non-idealities to give more accurate results.

After this study was concluded, $C_{010}$ was calculated for the known misalignment in the main HAYSTAC cavity, which was recently refurbished with a new axle. It was observed that the effects on the form factor were negligible, meaning the misalignment and resulting mode localization are at acceptable levels. The form factor for the perfectly aligned case ($\Delta f =0$) at approximately $90^\circ$ was 0.4654 whereas for a tilt of $\Delta f = 87.5$ kHz the form factor reduced to 0.4643, corresponding to a 0.24\% decrease. Similar calculations can be performed for any tuning position used in future data collection.

This study also revealed new insight on mode crossings. Mode crossings that were found to be small or benign in measurements showed the same behavior in simulations. In fact, simulations showed that in these cases the intruder mode in question was always a TEM mode. Previously, all mode crossings were viewed as unusable regions. For the existing cavity, these results demonstrate that some regions impacted by a mode crossing have minimal mixing, meaning portions of the mode crossing may still be useful for data collection. While mode crossings will always exist in any conventional cavity, future designs can be optimized to have large regions that are only interrupted by these small TEM mode crossings.

\section{Conclusions}\label{sec:conclusions}
This study has used microwave measurement techniques to characterize the HAYSTAC detector's resonant cavity in detail. The results will be used to inform how the cavity will be used in future axion search runs. Simulations of each measurement show that 3D electromagnetic models have the ability to capture the effects of non-idealities found in the cavity. There is great potential to continue to expand the model to capture the full complexity of the cavity. 

Further, future HAYSTAC cavities can be designed with more knowledge of the impact of potential fabrication errors and misalignments on cavity performance. By predicting the impacts of these flaws, designs can be optimized to maximize cavity performance even in non-ideal cases.

\begin{acknowledgments}
This work was supported by the National Science Foundation, under grants PHY-1362305 and PHY-1607417, and by the Heising-Simons Foundation under grants 2014-182 and 2016-044. We gratefully acknowledge the contributions of Al Kenany for his work on the vibration isolation system in the bead perturbation setup.
\end{acknowledgments}


%
%

%

\bibliography{cavitycharacterization}

\end{document}